\documentstyle[aps,prl,preprint,floats,epsfig]{revtex}

\begin{document}

\preprint{\tighten\vbox{\hbox{\hfil CLNS 02/1780}
                        \hbox{\hfil CLEO 02-03}
}}

\title{\large Anti-Search for the Glueball Candidate $f_J(2220)$ in Two-Photon Interactions}

\author{CLEO Collaboration}
\date{\today}

\maketitle
\tighten

\begin{abstract} 

Using 13.3 {\mbox{\rm fb$^{-1}$}} 
of $e^{+}e^{-}$ data recorded with the {CLEO\,II} and {CLEO\,II.V} detector configurations 
at CESR, we have searched  for {\mbox{$f_J(2220)$}}\ decays to {{\mbox{$K^0_{S}$}}{\mbox{$K^0_{S}$}}} 
in untagged two-photon interactions.
We report an upper limit on the product of the two-photon partial width and the branching 
fraction, $\Gamma_{{\mbox{$\gamma$}}{\mbox{$\gamma$}}}{\cal B}(f_J(2220){\mbox{$\rightarrow$}}{\mbox{$K^0_{S}$}}{\mbox{$K^0_{S}$}})$ 
of less than 1.1 eV at the 95$\%$ 
confidence level; systematic uncertainties are included. This dataset is
four times larger than that used in the previous CLEO publication.

\end{abstract}
\newpage

{
\renewcommand{\thefootnote}{\fnsymbol{footnote}}

 
\begin{center}
K.~Benslama,$^{1}$ B.~I.~Eisenstein,$^{1}$ J.~Ernst,$^{1}$
G.~D.~Gollin,$^{1}$ R.~M.~Hans,$^{1}$ I.~Karliner,$^{1}$
N.~Lowrey,$^{1}$ M.~A.~Marsh,$^{1}$ C.~Plager,$^{1}$
C.~Sedlack,$^{1}$
% ...
\end{center}
 
\small
\begin{center}
$^{1}${University of Illinois, Urbana-Champaign, Illinois 61801}\\
% ...
\end{center}

\setcounter{footnote}{0}
}
\newpage

 The $f_J(2220)$, also known as $\xi(2230)$, is a candidate for the lightest tensor
(J=2) 
glueball. The MARK-III Collaboration~\cite{markiii} first observed this state in 
the 
radiative 
decays $J/\psi${\mbox{$\rightarrow$}}{\mbox{$\gamma$}}{\mbox{$K^+$}}{\mbox{$K^-$}} and  
$J/\psi${\mbox{$\rightarrow$}}{\mbox{$\gamma$}}{\mbox{$K^0_{S}$}}{\mbox{$K^0_{S}$}}, in a sample of
$5.8\times 10^{6}$ $J/\psi$ decays. The masses (widths) of both modes
were consistent with those expected for a narrow tensor glueball; 
$2230.0\pm6.0\pm14.0~(26.0^{+20.0}_{-16.0}\pm17.0)$ {\mbox{\rm MeV/$c^2$}} and 
$2232.0\pm7.0\pm7.0~(18.0^{+23.0}_{-15.0}\pm10.0)$~{\mbox{\rm MeV/$c^2$}} for 
the {\mbox{$K^+$}}{\mbox{$K^-$}} and {\mbox{$K^0_{S}$}}{\mbox{$K^0_{S}$}} modes, 
respectively.  They did not observe any enhancement in the two-body 
final states {\mbox{$\pi^+$}}{\mbox{$\pi^-$}} and $p\bar{p}$. A year later, the DM2
Collaboration~\cite{dm2}, using a sample $8.6\times 10^{6}$ $J/\psi$ 
radiative decays, searched for the $f_J(2220)$ in the {\mbox{$\pi^+$}}
{\mbox{$\pi^-$}}, {\mbox{$K^+$}}{\mbox{$K^-$}},
and {\mbox{$K^0_{S}$}}{\mbox{$K^0_{S}$}} final states, and did not observe a signal in any of the three modes.
They reported a limit on the product branching fraction, 
${\cal B}(J/\psi{\mbox{$\rightarrow$}}{\mbox{$\gamma$}} f_J(2220)){\cal B}(f_J(2220)
{\mbox{$\rightarrow$}}{\mbox{$K^+$}}{\mbox{$K^-$}})$,
which was in disagreement with the MARK-III result. 
Ten years later, 
the
BES collaboration observed strong signals for $f_J(2220)$ decays into
{\mbox{$\pi^+$}}
{\mbox{$\pi^-$}}, {\mbox{$K^+$}}{\mbox{$K^-$}}, {\mbox{$K^0_{S}$}}{\mbox{$K^0_{S}$}}~\cite{bes1}, and 
{\mbox{$\pi^0$}}{\mbox{$\pi^0$}}~\cite{bes2}, 
again in radiative $J/\psi$ decays.
In hadron production the GAMS Collaboration~\cite{alde} reported a 
state, decaying to $\eta\eta^{\prime}$  in {\mbox{$\pi^-$}} p {\mbox{$\rightarrow$}}$\eta\eta^{\prime} n$
interactions, at 2220.0 {\mbox{\rm MeV/$c^2$}}. The angular distribution of the decay 
strongly indicated $J \geq 2$.  The LASS Collaboration~\cite{lass}
reported a narrow resonance, 
decaying to $K\bar{K}$.
Both the mass and the width of GAMS and LASS states were consistent with 
the previous $f_J(2220)$ measurements in radiative $J/\psi$ decays.

In 1997, the CLEO Collaboration reported tight limits on the two-photon
coupling of the $f_J(2220)$ in {\mbox{$\gamma$}}{\mbox{$\gamma$}}
{\mbox{$\rightarrow$}}{\mbox{$K^0_{S}$}}{\mbox{$K^0_{S}$}}~\cite{godang}
and {\mbox{$\gamma$}}{\mbox{$\gamma$}}{\mbox{$\rightarrow$}}{\mbox{$\pi^+$}}{\mbox{$\pi^-$}}~\cite{alam}.
Recent LEP results from the L3~\cite{lepl3} Collaboration showed no
evidence for $f_J(2220)$ production in two-photon interactions searching 
for the {\mbox{$K^0_{S}$}}{\mbox{$K^0_{S}$}}\ final state, and they derived an upper limit of  
$\Gamma_{{\mbox{$\gamma$}}{\mbox{$\gamma$}}}{\cal B}({\mbox{$f_J(2220)$}}{\mbox{$\rightarrow$}}
{\mbox{$K^0_{S}$}}{\mbox{$K^0_{S}$}}) 
< 1.4\ eV$ at 95\% confidence level (C.L.). 
under the hypothesis of a pure helicity-2 state.

Many experiments have searched for $f_J(2220)$ 
production
in $p\bar{p}$ 
annihilation-in-flight:
$p\bar{p}${\mbox{$\rightarrow$}}{\mbox{$\pi^+$}}{\mbox{$\pi^-$}}~\cite{hasan}, 
$p\bar{p}${\mbox{$\rightarrow$}}{\mbox{$K^+$}}{\mbox{$K^-$}}~\cite{bardin,sculli},
$p\bar{p}${\mbox{$\rightarrow$}}{\mbox{$K^0_{S}$}}{\mbox{$K^0_{S}$}}~\cite{evan1}, 
$p\bar{p}${\mbox{$\rightarrow$}}$\phi\phi$~\cite{evan2}, 
and $p\bar{p}{\mbox{$\rightarrow$}} p\bar{p}{\mbox{$\pi^+$}}{\mbox{$\pi^-$}}$~\cite{buzzo}.
None of the experiments have shown any evidence for a narrow $f_J(2220)$
resonance. Recent results of a high-sensitivity search in 
$p\bar{p}{\mbox{$\rightarrow$}}\eta\eta$,
$p\bar{p}{\mbox{$\rightarrow$}}\eta\eta^{\prime}$, and $p\bar{p}${\mbox{$\rightarrow$}}
{\mbox{$\pi^0$}}{\mbox{$\pi^0$}} reactions
from the Crystal Barrel Collaboration~\cite{seth} have also shown no evidence for 
the $f_J(2220)$ state.

The different experiments have shown contradictory results  
and it is clear that 
the existence and nature of the $f_J(2220)$ requires further experimental work. 
Here we report on a search of the {\mbox{$f_J(2220)$}}\ 
in untagged two-photon interactions at CLEO and a new upper limit on
the two-photon partial width times the branching fraction for its decay into 
{\mbox{$K^0_{S}$}}{\mbox{$K^0_{S}$}}.

 Color singlet hadronic states, such as mesons($q\bar{q}$) and baryons($qqq$) make 
bound states as a consequence of QCD color confinement. 
Color singlets can also be constructed
with gluons, hence named ``glueballs''. Glueballs are hadrons with no valence quarks,
bound together by the gluons' mutual attraction. Many different QCD-based models 
and calculations make predictions for such states: bag models~\cite{bar2,chan,detar},
constituent-glue models~\cite{horn,isgur1,isgur2}, QCD sum rules~\cite{lat}, 
and lattice 
gauge calculations based on the quenched approximation~\cite{cm,morn}.
Low mass scalar (J=0) glueballs are hard to 
detect and identify as their masses lie within the dense spectrum of conventional mesons, 
but heavy tensor glueballs are expected to be more easily observable. Gluons do 
not couple directly to 
photons (only via a box diagram)
and glueball two photon widths, $\Gamma_{{\mbox{$\gamma$}}{\mbox{$\gamma$}}}$, are expected to be small
relative to those of  mesons. A state that can readily be formed in a gluon rich 
environment, but not in two-photon collisions, has the quintessential signature of a glueball.
Upper limits derived from {\mbox{$\gamma$}}{\mbox{$\gamma$}}\ data can thus play a major role
in identifying glueballs.

 The data presented here were taken by the {CLEO\,II}~\cite{kubota} 
and {CLEO\,II.V}~\cite{hill} detector 
configurations operating at the Cornell Electron Storage Ring. The sample used in this analysis 
corresponds to an integrated 
$e^{+}e^{-}$
luminosity of 13.3 {\mbox{\rm fb$^{-1}$}} from data taken on the {\mbox{$\Upsilon(4S)$}} and at the 
energies just below the {\mbox{$\Upsilon(4S)$}}. This is four times the sample size used in the
prior CLEO publications\cite{godang,alam}.  The CLEO detector 
includes several concentric tracking
devices to detect and measure charged particles over 95$\%$ of 
4$\pi$ steradians and a CsI electromagnetic calorimeter, both 
operating inside 1.5 T superconducting solenoid.
The tracking system in {CLEO\,II}~\cite{kubota} consisted of a 6-layer straw tube chamber, a
10-layer precision tracker and a 51-layer main drift chamber.
For {CLEO\,II.V}~\cite{hill} the straw tube chamber was replaced 
by a 3-layer, double-sided silicon vertex detector, and the gas in the main drift chamber was 
changed from an argon-ethane to a helium-propane mixture. This change
in gas improved both the hit efficiency and the specific ionization 
resolution\cite{peterson}.

The Monte Carlo generation of two photon production is modeled on the BGMS
formalism~\cite{bgms}, for which we assumed {\it J}\,=\,2 for the 
glueball candidate.
The simulation of the transport and decay of the final state particles through the CLEO detector 
is performed by the GEANT package~\cite{geant}. We estimate a {\mbox{$K^0_{S}$}}{\mbox{$K^0_{S}$}}\ 
mass resolution of $\sigma=$ 7.86 
{\mbox{\rm MeV/$c^2$}} 
near the  PDG average mass\cite{pdg} of 2231 {\mbox{\rm MeV/$c^2$}}. 
The net efficiencies for the 0 and $\pm$2 helicities
are 13.6$\%$ and 19.1$\%$, respectively; this includes the 69\% branching fraction
for each $K_{S}^{0}\to\pi^{+}\pi^{-}$.

\begin{figure}[h]
\begin{center}
\hspace{-0.1in}\mbox{\psfig{figure=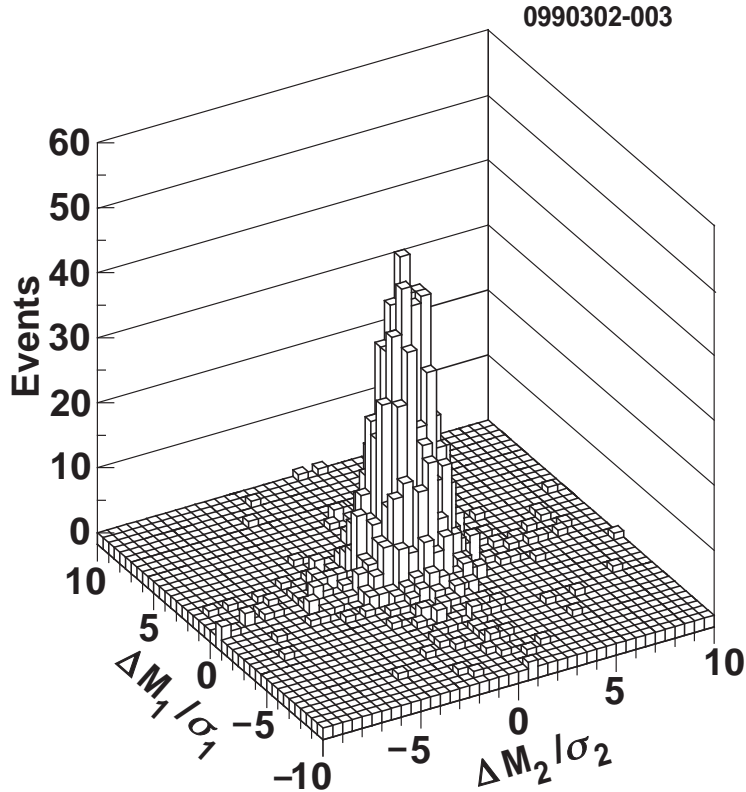,width=4.0in}}
\caption{
$\Delta M_{1}/\sigma_{1}~vs.~\Delta M_{2}/\sigma_{2}$
for data, with $\Delta M$ being the difference between the invariant mass
of a dipion combination and the known $K^{0}_{S}$ mass. 
We select good {\mbox{$K^0_{S}$}}{\mbox{$K^0_{S}$}}\ 
candidates within a circle of radius 3.5 units. }
\label{fig:ksks}
\end{center}
\end{figure}

The kinematics of untagged two-photon events are defined by the fact that the photons have
a large fraction of their momenta along the beam line; that is the two photons
are almost on mass shell. The scattered electron 
and positron do not in general have sufficient transverse momentum to be 
detected in the tracking 
chambers. The two photons rarely have the same magnitude of momentum, and as a result 
the two-photon center of mass is boosted along the beam axis. We select events 
containing exactly four reconstructed charged tracks with zero net charge.
To ensure these events have no accompanying photon showers,
we require the unmatched neutral energy to be less than 0.6 {\rm GeV}. To suppress non-two photon 
events ($udsc$ continuum and $\tau^{+}\tau^{-}$ events) 
we require the total charged track energy be less than 4.5 {\rm GeV} and that the
vector sum of transverse momentum of all charged tracks be 
less than 0.6 {\mbox{\rm GeV/$c$}}~ in magnitude. To suppress 
two-photon events that do not have $K_{S}^{0}$ mesons in the final state,
we applied a flight distance significance criterion (flight distance divided by 
its uncertainty)
of 3 for {CLEO\,II} and 5 for {CLEO\,II.V} data, 
and required that each charged pion daughter of the {\mbox{$K^0_{S}$}}\ 
candidates not point 
back to the interaction point. Finally, we selected 
good events with two
good {\mbox{$K^0_{S}$}}\ candidates by requiring  
($\Delta M_{1}/\sigma_{1},\Delta M_{2}/\sigma_{2}$) 
to lie within a circle of radius 3.5.
Here $\Delta M = m_{\pi\pi} - m_{K}$,
$m_{K}$ is the $K_{S}^{0}$ mass of 497.7 MeV/$c^{2}$,
and $\sigma_1$ and $\sigma_2$ are the mass resolutions for the two {\mbox{$K^0_{S}$}}\ 
candidates calculated on an event-by-event basis. 
In Fig.~\ref{fig:ksks} we show the 
distribution of these scaled mass differences within 
$\pm$10 $\sigma$ of the nominal {\mbox{$K^0_{S}$}}\  mass.
We conclude from Fig.~\ref{fig:ksks} that we have no substantial background
that does not contain $K^{0}_{S}$ mesons. 

\begin{figure}[h]
\begin{center}
\hspace{-0.1in}\mbox{\psfig{figure=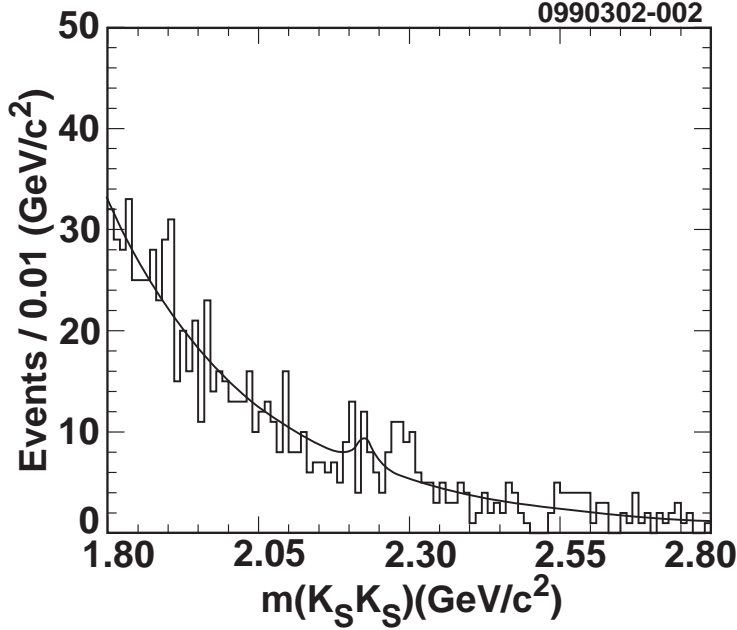,width=4.0in}}
\caption{ {\mbox{$K^0_{S}$}}{\mbox{$K^0_{S}$}}\ mass distribution observed in data around {\mbox{$f_J(2220)$}}\ mass
region. The solid line is the sum of a fit to the background and the
signal line shape which was obtained from Monte Carlo. The number of observed
events at 95$\%$ C.L. upper limit is 29.9.}
\label{fig:fj}
\end{center}
\end{figure}

 Using the data sample described above, we combine the two {\mbox{$K^0_{S}$}}\ 
candidates in the event  
and plot the invariant mass distribution (Fig.~\ref{fig:fj}) from 1.8 to 2.8 {\mbox{\rm GeV/$c^2$}}. 
We fit the data to the combination of a power law 
background function ($A W_{{\mbox{$\gamma$}}{\mbox{$\gamma$}}}^n$) 
and a signal shape comprising a Breit-Wigner convolved with a Gaussian 
resolution function derived from the
Monte Carlo studies. The mass and Breit-Wigner width of the signal function are 
allowed to float 
within $\pm 1 \sigma$ of the PDG~\cite{pdg} values for the
$f_J(2220)$: mass of 2231$\pm$3.5 {\mbox{\rm MeV/$c^2$}} 
and width of $23^{+7}_{-8}$ {\mbox{\rm MeV/$c^2$}}.
A statistically insignificant excess of $15\pm11$ events is found in the signal 
region, with a mass of 2228 MeV/$c^{2}$ and a width of 31 MeV/$c^{2}$, corresponding to
an upper limit of 29.9 events at the 95$\%$ confidence level. The largest excess 
apparent in the plot is 
one of 24$\pm$8 events at a mass of 2290.0 {\mbox{\rm MeV/$c^2$}}. There are no 
known narrow resonances in this region, 
and we consider this enhancement to be a statistical fluctuation.
The mass of 2290.0 {\mbox{\rm MeV/$c^2$}} for this excess is also completely inconsistent 
with the previous measurements of the $f_J(2220)$. Allowing this enhancement 
into the fit 
(describing it with a single Gaussian with its width allowed to float)
slightly lowers 
the excess found in the 2231 {\mbox{\rm MeV/$c^2$}} region. We therefore conservatively
quote our result without this excess at 2290.0 {\mbox{\rm MeV/$c^2$}} in the fit.
In this analysis we do 
not have 
enough events in the high mass region to interpret quantitatively any interference
effect between resonant ({\mbox{$\gamma$}}{\mbox{$\gamma$}}
{\mbox{$\rightarrow$}}{\mbox{$f_J(2220)$}}{\mbox{$\rightarrow$}}{\mbox{$K^0_{S}$}}{\mbox{$K^0_{S}$}}) 
and non-resonant ({\mbox{$\gamma$}}{\mbox{$\gamma$}}{\mbox{$\rightarrow$}}{\mbox{$K^0_{S}$}}{\mbox{$K^0_{S}$}}) events.
Therefore in the above fit we did not include an interference term between them.

 To extract the value of 
$\Gamma_{{\mbox{$\gamma$}}{\mbox{$\gamma$}}}{\cal B}(f_{J}\to K^{0}_{S}K^{0}_{S})$ 
for {\mbox{$f_J(2220)$}}~ from the data, we scaled 
the branching fraction and the partial width used in the Monte Carlo 
production by the ratio of 
the upper limit on the number of data events ($n^{\rm data}$) to the number of 
Monte Carlo events passing our selection criteria ($n^{\rm MC}$),
and the ratio of Monte Carlo to data luminosities,

\begin{equation}
\Gamma_{{\mbox{$\gamma$}}{\mbox{$\gamma$}}}{\cal B}(f_{J}\to K^{0}_{S}K^{0}_{S})=
\frac{n^{data}}{n^{\rm MC}} \frac{{\cal L}^{\rm MC}}{{\cal L}^{\rm data}}
[\Gamma_{{\mbox{$\gamma$}}{\mbox{$\gamma$}}}{\cal B}(f_{J}\to K^{0}_{S}K^{0}_{S})]^{\rm MC}.
\end{equation}           

Note that this procedure is independent of the actual values used
for $\Gamma_{{\mbox{$\gamma$}}{\mbox{$\gamma$}}}$ and ${\cal B}(f_{J}\to K^{0}_{S}K^{0}_{S})$
in the simulation (which were 1 keV and 1.0, respectively).

  Our estimate of the systematic uncertainties in the 
overall detector efficiency is: 
7\% due to event selection criteria,
5\% due to trigger effects, 4\% due to tracking, 
3\% from online software filtering.
We assign systematic uncertainties of
8\% from our background parameterization and 1\% due 
to the luminosity measurement. We add 
these in quadrature to obtain a total systematic uncertainty of 13\%.

 Spin 2 resonances from two-photon events can have two helicity projections, 
0 and 2. $\Gamma_{{\mbox{$\gamma$}}{\mbox{$\gamma$}}}$ is therefore 
a superposition of two components, $\Gamma^{2,0}_{{\mbox{$\gamma$}}{\mbox{$\gamma$}}}$ 
and $\Gamma^{2,2}_{{\mbox{$\gamma$}}{\mbox{$\gamma$}}}$. The efficiencies for these 
two helicities, given above,  are different due to their 
different final state angular distributions. 
States with helicity of 2 should dominate those of helicity zero
in a 6:1 ratio, based purely on the Clebsch-Gordon coefficients.
The number of Monte Carlo events observed and the Monte Carlo 
luminosities were therefore
reweighted to this 6:1 ratio~\cite{poppe,kolan}, for the $|Y^2_2|^2$ and  
$|Y^0_2|^2$ angular distributions, yielding
\begin{equation}
\Gamma_{{\mbox{$\gamma$}}{\mbox{$\gamma$}}}{\cal B}(f_{J}\to K^{0}_{S}K^{0}_{S}) 
\leq 1.1\,eV,~~~~(95\,\%~C.L.)
\end{equation}
This limit includes our  systematic uncertainties. 
Alternatively, we also present 
our results as a functional limit for a state with 
{\it J} = 2, without assuming the ratio of partial widths of 
the two helicity projections,

\begin{equation}
(0.53\Gamma^{2,0}_{{\mbox{$\gamma$}}{\mbox{$\gamma$}}}~+~1.08
\Gamma^{2,2}_{{\mbox{$\gamma$}}{\mbox{$\gamma$}}}) {\cal B}(f_{J}\to K^{0}_{S}K^{0}_{S})
\leq 1.1\,eV ~~~~(95\,\%~C.L.)
\end{equation}
The ratio of the coefficients is the ratio of the efficiencies 
for the two helicities, normalized 
to the 6:1 ratio in Eq.(2).

 To build confidence in our approach for the {\mbox{$f_J(2220)$}}\ search, we have also checked
the two-photon partial width and the mass
of the well established {\mbox{$f_2^{\prime}(1525)$}}\ resonance using the same Monte Carlo 
simulation and analysis technique, and find the
values for both the width and the mass consistent with the PDG values~\cite{f2prime}.

Under the assumption that the $f_{J}$ resonance has a large branching fraction
to kaons,  the low limit on the 
$\Gamma_{{\mbox{$\gamma$}}{\mbox{$\gamma$}}}{\cal B}(f_{J}\to K^{0}_{S}K^{0}_{S})$ 
implies 
that the {\mbox{$f_J(2220)$}}\ production is 
very suppressed in two-photon collisions. 
This is exactly the behavior that would be expected 
for a true glueball as gluons, being neutral, do not 
couple with electric charge.
Quantitatively,
a naive glueball figure of merit known as ``stickiness'' is frequently used. 
Stickiness is a measure of color charge 
relative to electric charge\cite{cha}:
\begin{equation}
S_X~=~N_l \left(\frac{m_X}{k_{X}}\right)^{2l+1}
\frac{\Gamma(\psi{\mbox{$\rightarrow$}}
{\mbox{$\gamma$}} X)}{\Gamma(X{\mbox{$\rightarrow$}}{\mbox{$\gamma$}}{\mbox{$\gamma$}})}.  
\end{equation}    
Where, $k_{X}~=~(m^2_{\psi}\,-\,m^2_X)/2m_{\psi}$ is the energy
of the photon from the radiative $J/\psi$ decay in the $J/\psi$ rest frame.
$N_l$ is the normalization factor and is so chosen that $S_X\,=\,1$
for the {\mbox{$f_2(1270)$}}\ meson.   

In order to determine an upper limit for stickiness, we first extract
average the results of the 
MARK\,III~\cite{markiii} and BES~\cite{bes1} experiments, 
obtaining 
${\cal B}(J/\psi{\mbox{$\rightarrow$}}{\mbox{$\gamma$}} f_J(2220))
{\cal B}(f_J(2220){\mbox{$\rightarrow$}}{\mbox{$K^0_{S}$}}{\mbox{$K^0_{S}$}})$ = $(2.2\pm0.6)\times 10^{-5}$.
Within each 
experiment we form  a branching fraction to $K\bar{K}$ by adding 
twice the branching fraction to {\mbox{$K^0_{S}$}}{\mbox{$K^0_{S}$}} to that for {\mbox{$K^+$}}{\mbox{$K^-$}}. 
We combine our upper 
limit for 
$\Gamma_{{\mbox{$\gamma$}}{\mbox{$\gamma$}}}{\cal B}(f_{J}\to K^{0}_{S}K^{0}_{S})$ 
with this  
product branching fraction 
and the $J/\psi$  width of $\Gamma = 87$ keV\cite{pdg} to
set a lower limit on the value of the $f_{J}(2220)$ ``stickiness'' 
of 109 at the 95$\%$ C.L.

In conclusion, we do not see a signal for {\mbox{$f_J(2220)$}}\ in the {\mbox{$K^0_{S}$}}{\mbox{$K^0_{S}$}}\ invariant mass
distribution at masses near those reported by previous experiments\cite{pdg}. 
Therefore, we set an upper limit on 
$\Gamma_{{\mbox{$\gamma$}}{\mbox{$\gamma$}}}
{\cal B}(f_{J}{\mbox{$\rightarrow$}}{\mbox{$K^0_{S}$}}{\mbox{$K^0_{S}$}})$ for the 
$f_J(2220)$ of 1.1 eV at the 95\% confidence level. 
We allowed the width and mass to float
within $\pm 1$ $\sigma$ of the PDG values. 
Our limit is lower than the previous CLEO 
measurement based on a quarter of the present luminosity. Recently, 
L3~\cite{lepl3} published an
upper limit for 
$\Gamma_{{\mbox{$\gamma$}}{\mbox{$\gamma$}}}{\cal B}(f_{J}(2220){\mbox{$\rightarrow$}}
{\mbox{$K^0_{S}$}}{\mbox{$K^0_{S}$}})$ of  1.4 eV
at 95\% C.L., which is similar to our upper limit of 1.02 eV 
(from Eq. 3) at 95\% C.L.
under the hypothesis of a pure helicity-2 state. 
This low value of 
$\Gamma_{{\mbox{$\gamma$}}{\mbox{$\gamma$}}}
{\cal B}(f_{J}(2220){\mbox{$\rightarrow$}}{\mbox{$K^0_{S}$}}{\mbox{$K^0_{S}$}})$ 
indicates that the 
$f_J(2220)$ coupling to photons is suppressed and argues the case that, 
if the previous observations of the $f_J(2220)$ by the MARK III and BES collaborations
in radiative $J/\psi$ decays
are correct, it has the signature of a glueball. On the other hand, our data is also 
consistent with the non-existance of any narrow resonance in the mass region.

We gratefully acknowledge the effort of the CESR staff in providing us with
excellent luminosity and running conditions.
M. Selen thanks the PFF program of the NSF and the Research Corporation,
and A.H. Mahmood thanks the Texas Advanced Research Program.
This work was supported by the National Science Foundation, and the
U.S. Department of Energy.


\begin{thebibliography}{99}
\bibitem{markiii}
MARK-III Collaboration, R.M. Baltrusaitis {\em et al.}, Phys. Rev. Lett. {\bf 56}, 107 (1986).
\bibitem{dm2}
J. E. Augustin {\em et al.}, Z. Phys. C {\bf 36}, 369 (1987).
\bibitem{bes1}
BES Collaboration, J.Z. Bai {\em et al.}, Phys. Rev. Lett. {\bf 76}, 3502 (1996).
\bibitem{bes2}
BES Collaboration, J.Z. Bai {\em et al.}, Phys. Rev. Lett. {\bf 81}, 1179 (1998).
\bibitem{alde}
GAMS Collaboration, D. Alde {\em et al.}, Phys. Lett. B {\bf 177}, 120 (1986).
\bibitem{lass}
LASS Collaboration, D. Aston {\em et al.}, Phys. Lett. B {\bf 215}, 199 (1988).
\bibitem{godang}
CLEO Collaboration, R. Godang {\em et al.}, Phys. Rev. Lett. {\bf 79}, 3829 (1997).
\bibitem{alam}
CLEO Collaboration, M.S. Alam {\em et al.}, Phys. Rev. Lett. {\bf 81}, 3328 (1998).
\bibitem{lepl3}
L3 Collaboration, M. Accirri {\em et al.}, Phys. Lett. B {\bf 501}, 173 (2001).
\bibitem{hasan}
A. Hasan and D.V. Bugg, Phys. Lett. B {\bf 388}, 376 (1996).
\bibitem{bardin}
G. Bardin {\em et al.}, Phys. Lett. B {\bf 195}, 292 (1987).
\bibitem{sculli}
J. Sculli, J. H. Christenson, G. A. Kreiter, P. Nemethy and P.Yamin,
Phys. Rev. Lett. {\bf 58}, 1715 (1987).
\bibitem{evan1}
Jetset Collaboration, C. Evangelista {\em et al.}, Phys. Rev. D {\bf 56}, 3803 (1997). 
\bibitem{evan2}
Jetset Collaboration, C. Evangelista {\em et al.}, Phys. Rev. D {\bf 57}, 5370 (1998). 
\bibitem{buzzo}
Jetset Collaboration, A. Buzzo {\em et al.}, Z. Phys. C {\bf 76}, 475 (1997).
\bibitem{seth}
Crystal Barrel Collaboration, K.K.~Seth, Nuc.\ Phys.\ {\bf A663}, 600 (2000).
\bibitem{bar2}
T. Barnes, F.E. Close, and F. de Viron, Nuc. Phys.\ {\bf B224}, 241 (1983).
\bibitem{chan}
M.S. Chanowitz and S. Sharpe, Phys. Lett. B {\bf 132}, 413 (1983).
\bibitem{detar}
C. DeTar and J. Donoghue, Annu. Rev. Nucl. Part. Sci. {\bf 33}, 325 (1983).
\bibitem{horn}
D. Horn and J. Mandula, Phys. Rev. D {\bf 17}, 898 (1978).
\bibitem{isgur1}
N. Isgur and J. Paton, Phys. Lett. B {\bf 124}, 247 (1983);
N. Isgur and J. Paton, Phys. Rev. D {\bf 31}, 2910 (1985).

\bibitem{isgur2}
N. Isgur, R. Kokoski, and J. Paton, Phys. Rev. Lett. {\bf 54}, 869 (1985).

\bibitem{lat}
J. L. Latorre, I. S. Narison, P. Pascual, and R. Tarrach, Phys. Lett. B {\bf 147}, 169 (1984).

\bibitem{cm}
C. Michael and M. Teper, Nuc. Phys.\ {\bf B314}, 347 (1989).

\bibitem{morn}
C.J. Morningstar and M. Peardon, Phys. Rev. D {\bf 60}, 034509 (1999).

\bibitem{kubota}
CLEO Collaboration, Y. Kubota {\em et al.}, Nucl. Instr. and Meth. A {\bf 320}, 66 (1992).

\bibitem{hill}
T. Hill, Nucl. Instr. and Meth. A {\bf 418}, 32 (1998).

\bibitem{peterson}
D. Peterson, Nuc. Phys. B (Proc. Suppl), {\bf 54B}, 31 (1997).

\bibitem{bgms}
V. M. Budnev, I. F. Ginzburg, G. V. Meledin, and V. G. Serbo, Phys. Rep.
C {\bf 15}, 181 (1996).

\bibitem{geant}
R. Brun {\em et al.}, ``GEANT, Detector Description and Simulation Tool,''
CERN Program  Library Long Writeup W5013, 1993.
\bibitem{pdg}
Particle Data Group, D.E. Groom {\em et al.}, Eur. Phys. J. C {\bf 15}, 1 (2000).
\bibitem{poppe}
M. Poppe, Int. J. Mod. Phys, A {\bf 1}, 545 (1986).
\bibitem{kolan}
H. Kolanoski and P. Zerwas, in {\it High Energy Electron-Positron
Physics}, edited by A.~Ali and P.~Soding (World Scientific, Singapore, 1988), p. 695. 
\bibitem{f2prime}
We found the two-photon partial width of the {\mbox{$f_2^{\prime}(1525)$}}\ to be 
$91.3\pm 5.4$ eV, where the error is statistical only. This 
may be compared with the PDG value of
97$\pm$16 eV~\cite{pdg}. The PDG mass for {\mbox{$f_2^{\prime}(1525)$}}\ 
is 1524.6$\pm$1.4 {\mbox{\rm MeV/$c^2$}}.
\bibitem{cha}
M.S. Chanowitz, 1984, in {\it Proceedings of the VI International Workshop
on Photon-Photon Collisions, Lake Tahoe, CA}, edited by R. L. Lander
(World Scientific, Singapore, 1984), p. 95.        
\end{thebibliography}
\end{document}